\documentclass{aa}  

\usepackage{graphicx}
\usepackage{wasysym}
\usepackage{txfonts}
\usepackage{soul}
\newcommand{\change}[1]{#1}          
\newcommand{\delete}[1]{}                

\newcommand{\changetwo}[1]{#1}          
\newcommand{\deletetwo}[1]{}                

\newcommand{\changethree}[1]{#1}          
\newcommand{\deletethree}[1]{}                

\begin{document}

   \title{Evidence for depletion of heavy silicon isotopes at comet 67P/Churyumov-Gerasimenko}


   \author{M. Rubin
          \inst{1}\fnmsep\thanks{\email{martin.rubin@space.unibe.ch}}
          \and
          K. Altwegg\inst{1,2}
          \and
          H. Balsiger\inst{1}
          \and
          J.-J. Berthelier\inst{3}
          \and
          A. Bieler\inst{1,4}
          \and
          U. Calmonte\inst{1}
          \and
          M. Combi\inst{4}
          \and
          J. De Keyser\inst{5}
          \and
          C. Engrand\inst{6}
          \and
          B. Fiethe\inst{7}
          \and
          S. A. Fuselier\inst{8,9}
          \and
          S. Gasc\inst{1}
          \and
          T. I. Gombosi\inst{4}
          \and
          K. C. Hansen\inst{4}
          \and
          M. H\"assig\inst{1,8}
          \and
          L. Le Roy\inst{1}
          \and
          K. Mezger\inst{2,10}
          \and
          C.-Y. Tzou\inst{1}
          \and
          S. F. Wampfler\inst{2}
          \and
          P. Wurz\inst{1,2}
          }

   \institute{Physikalisches Institut, University of Bern, Sidlerstrasse 5, CH-3012 Bern, Switzerland
         \and
                Center for Space and Habitability, University of Bern, Sidlerstrasse 5, CH-3012 Bern, Switzerland
         \and
         LATMOS 4 Avenue de Neptune F-94100 SAINT-MAUR, France
         \and
         Climate and Space Sciences and Engineering, University of Michigan, Ann Arbor, MI 48109, USA
         \and
         Royal Belgian Institute for Space Aeronomy (BIRA-IASB), Ringlaan 3, B-1180, Brussels, Belgium
         \and
         Centre de Sciences Nucl\'eaires et de Sciences de la Mati\`ere, CNRS/IN2P3, Universit\'e Paris Sud, UMR 8609, Universit\'e Paris-Saclay, 91405 Orsay Campus, France
         \and
         Institute of Computer and Network Engineering (IDA), TU Braunschweig, Hans-Sommer-Strasse 66, D-38106 Braunschweig, Germany
         \and
         Space Science Directorate, Southwest Research Institute, 6220 Culebra Road, San Antonio, TX 78228, USA
         \and
        Department of Physics and Astronomy, University of Texas at San Antonio, San Antonio, TX 78249, USA
         \and
         Institut f\"ur Geologie, University of Bern, Baltzerstrasse 1+3, CH-3012 Bern, Switzerland
             }


 
  \abstract
   {The Rosetta Orbiter Spectrometer for Ion and Neutral Analysis (ROSINA) was designed to measure the composition of the gas in the coma of comet 67P/Churyumov-Gerasimenko, the target of the European Space Agency's Rosetta mission. In addition to the volatiles, ROSINA measured refractories sputtered off the comet \delete{ due to}\change{by} the interaction of solar wind protons with the surface of the comet.}
   {The origin of different solar system materials is still heavily debated. Isotopic ratios can be used to distinguish between different reservoirs and investigate processes occurring during the formation of the solar system.}
   {ROSINA consisted of two mass spectrometers and a pressure sensor. In the ROSINA Double Focusing Mass Spectrometer (DFMS), the neutral gas of cometary origin was ionized and then deflected in an electric and a magnetic field that separated the ions based on their mass-to-charge ratio. The DFMS had a high mass resolution, dynamic range, and sensitivity that allowed detection of rare species and the known major volatiles.}
   {We measured the relative abundance of all three stable silicon isotopes with the ROSINA instrument on board the Rosetta spacecraft. Furthermore, we measured $^{13}$C/$^{12}$C in C$_2$H$_4$, C$_2$H$_5$, and CO. The DFMS in situ measurements \change{indicate} that the average silicon isotopic composition \delete{is depleted}\change{shows depletion} in the heavy isotopes $^{29}$Si and $^{30}$Si with respect to $^{28}$Si and solar abundances, while  $^{13}$C to $^{12}$C \change{is analytically indistinguishable from bulk planetary and meteorite compositions.} \delete{corresponds to bulk planetary and meteoritic C within error bars. While}\change{Although} the origin of the deficiency of the heavy silicon isotopes cannot be explained unambiguously, we discuss mechanisms that could \change{have} contribute\change{d} to the measured depletion of the \changethree{isotopes} $^{29}$Si and $^{30}$Si.\deletethree{  isotopes."silicon isotopes" would not be hyphenated, so I have removed the hyphen here and throughout. Here I think it should read "... of the isotopes $^{29}$Si and $^{30}$Si" , OK}}
   {}
   

   \maketitle
%

\section{Introduction}

In August 2014 the European Space Agency's Rosetta spacecraft arrived at comet 67P/Churyumov-Gerasimenko (hereafter 67P) and started an in-depth investigation of the comet and its surrounding coma \citep{RN378}. One of the main goals early in the mission was to find a suitable landing site for the Philae lander module, which was then successfully deployed on 12 November 2014 on the surface of the comet. For this purpose the Rosetta spacecraft performed ever-closer orbits around the comet. This provided the Rosetta Orbiter Spectrometer for Ion and Neutral Analysis (ROSINA) with optimal observing conditions as the neutral densities increased with decreasing distances to the nucleus.

ROSINA was part of the Rosetta orbiter science payload and measured the different volatile\delete{s} \change{elements and compounds} and their abundances in the coma of the comet using  the Double Focusing Mass Spectrometer (DFMS) \citep{RN13}. The DFMS performed continuous measurements in the coma including an orbital arc from the end of October to the beginning of November 2014 over the southern winter hemisphere of the comet. At this early stage in the mission at a heliocentric distance of 3.07 au, the comet's coma was still thin over the poorly sunlit southern hemisphere, allowing at least some solar wind protons to hit the surface unhindered. This period therefore proved to be fruitful for measuring atoms sputtered from the surface by impinging solar wind protons. \cite{RN561} reported relative abundances of $^{23}$Na, $^{28}$Si, $^{39}$K, and $^{40}$Ca isotopes of which $^{28}$Si was the most abundant; however, the average elemental ratios found did not match the bulk composition of any specific type of known chondritic material. In this paper, we follow up the constraints from the elemental composition with measurements of the isotopic composition of Si.

Silicon isotopes can be used to study the processes that shaped the refractory material before it was incorporated into the comet. It has been shown that elements heavier than C are formed in stars from which they are expelled during disruptive events such as supernova explosions or through stellar winds \citep{RN649}. It is therefore the stellar atmosphere that defines the composition of the grain\delete{ once it condenses}. The \change{composition of the} atmosphere, in turn, is \delete{defined}\change{controlled} by the matter from which the star originally formed \delete{in combination with}\change{and by the} nucleosynthetic processes and pathways \citep[e.g.,][]{RN656}. The interstellar medium (ISM) is a collection of these materials and represents the \delete{origin}\change{primary source} of the matter in planetary systems\delete{ such as}\change{, including} our solar system. For this reason, relative abundances of elements and their isotopes not only give insight into the processes inside the star and possibly its catastrophic end, but also into the evolution of the ISM and the protosolar nebula through the decay of long-lived radio nuclides, the interaction with cosmic rays, and mass-dependent as well as independent fractionation processes \citep{RN655}.

The observation of heterogeneous isotopic abundances in primitive meteorites reveals details about the chemical evolution and the corresponding physical environment showing that the material in our solar system originates from different sources and is only partially mixed\deletethree{, not fully mixed} and \changethree{not fully} homogenized. In particular, presolar grains such as silicon carbide (SiC) and graphite exhibit distinct isotopic ratios in many elements compared to the bulk solar system ratios \change{\citep{zinner1998stellar, nittler2003presolar, davis2011stardust}}. In contrast, the ratios observed in calcium-aluminum-rich inclusions (CAIs), which formed early in our solar system, show much smaller deviations with respect to the bulk material of the solar system. Through the analysis of primitive meteorites, the heterogeneity of presolar grains is well established. They are mostly observed in the form of diamonds, silicates, oxides, SiC, graphite, and silicon nitride, but constitute only a small fraction of the mass and are embedded in matter processed during the formation of our solar system \citep{RN656, hoppe2009stardust, floss2016presolar}. These characteristics show that presolar material in part survives the collapse of the protosolar nebula and the formation of the Sun and the planets. It seems clear, however, that the carbonaceous interstellar grains found in meteorites do not represent the bulk presolar matter, i.e., they themselves are special because  they survived not only the incorporation into the meteorite at high temperatures, but also the oxidizing environment of the solar system \citep{RN655}. The presolar origin of a portion  of (or the precursor to) the organic matter in primitive chondrites is also debated \citep[e.g.,][]{alexander2010deuterium, alexander2014elemental, remusat2016thermal}.

Comets are thought to represent an even more pristine component of solar system material than meteorites. The latest results from the Rosetta mission also indicate that the nucleus of the comet itself is primordial and still contains layered features obtained during the formation process \citep{RN598}, although this is contested by current solar system formation models \citep{JutziColl}. However, neither theory contradicts the supposition that parts of the ices could be of interstellar origin and never sublimated. This is supported by the elevated HDO/H$_2$O ratio in the water of 67P  \citep{RN462}, which -- according to recent modeling results \citep{RN654} -- cannot be produced in a protoplanetary disk (and requires a significant amount of water ice inherited from the earliest stages of star formation). Further evidence comes from an even higher D$_2$O/HDO ratio \citep{AltweggD2O}, \delete{comparable}\change{similar} to interstellar ratios \citep{1538-4357-659-2-L137,2041-8205-792-1-L5}. This conclusion is consistent with the detection of highly volatile species incorporated in the cometary ices \citep{RN671}, such as  CH$_4$ and CO \citep{RN350,RN592}, N$_2$ \citep{RN505}, Ar \citep{RN596}, and the amounts of O$_2$ that  correlate well with H$_2$O \citep{RN665}. The most recent observation of volatile molecular sulfur, S$_2$, even suggests that sublimation and re-condensation with the trapping of volatiles becomes improbable, due to the very short photo-lifetime of S$_2$ \citep{Calmonte08102016}. Given the strong indication that at least parts of the ices in 67P are older than our solar system, the focus is again on the refractory material. From what we know so far, the cometary dust and ice is well mixed and no large-scale heterogeneities seem to be present \citep{RN653}, which is supported by CONSERT measurements through the smaller lobe \citep{RN556}. However, the degree of mixing of ices and refractory material is not fully understood. Isotopic ratios should give some clues about the distribution and the transport of the refractories in the early solar system. This includes the possibilities that the material in 67P is of presolar origin or formed closer to the Sun and then was transported later, for example through radial mixing, to the comet formation region before incorporation with the unaltered ices. For this purpose we analyzed measurements obtained by ROSINA DFMS shortly before lander delivery in November 2014. The mass spectra show the presence of the three stable Si isotopes and allow a first look at the isotopic ratio of Si sputtered by the solar wind off the dust particles in this comet. No higher mass molecule containing Si such as SiH or SiO has been identified, which supports the theory that Si was sputtered as an atom from the surface. For reference, we compare our data to the relative solar abundances of the three stable Si isotopes from \cite{RN664} of 92.230\%, 4.683\%, and 3.087\% for $^{28}$Si, $^{29}$Si, and $^{30}$Si ($^{29}$Si/$^{28}$Si=0.0508 or $^{28}$Si/$^{29}$Si=19.7 and $^{30}$Si/$^{28}$Si=0.0335 or $^{28}$Si/$^{30}$Si=29.9), respectively, which all fall within the ranges provided by the Commission on Isotopic Abundances and Atomic Weights of the International Union of Pure and Applied Chemistry \citep{berglund2011isotopic}.

\section{ROSINA DFMS sensor on Rosetta}

   \begin{figure*}
   \centering
   \includegraphics[width=150mm]{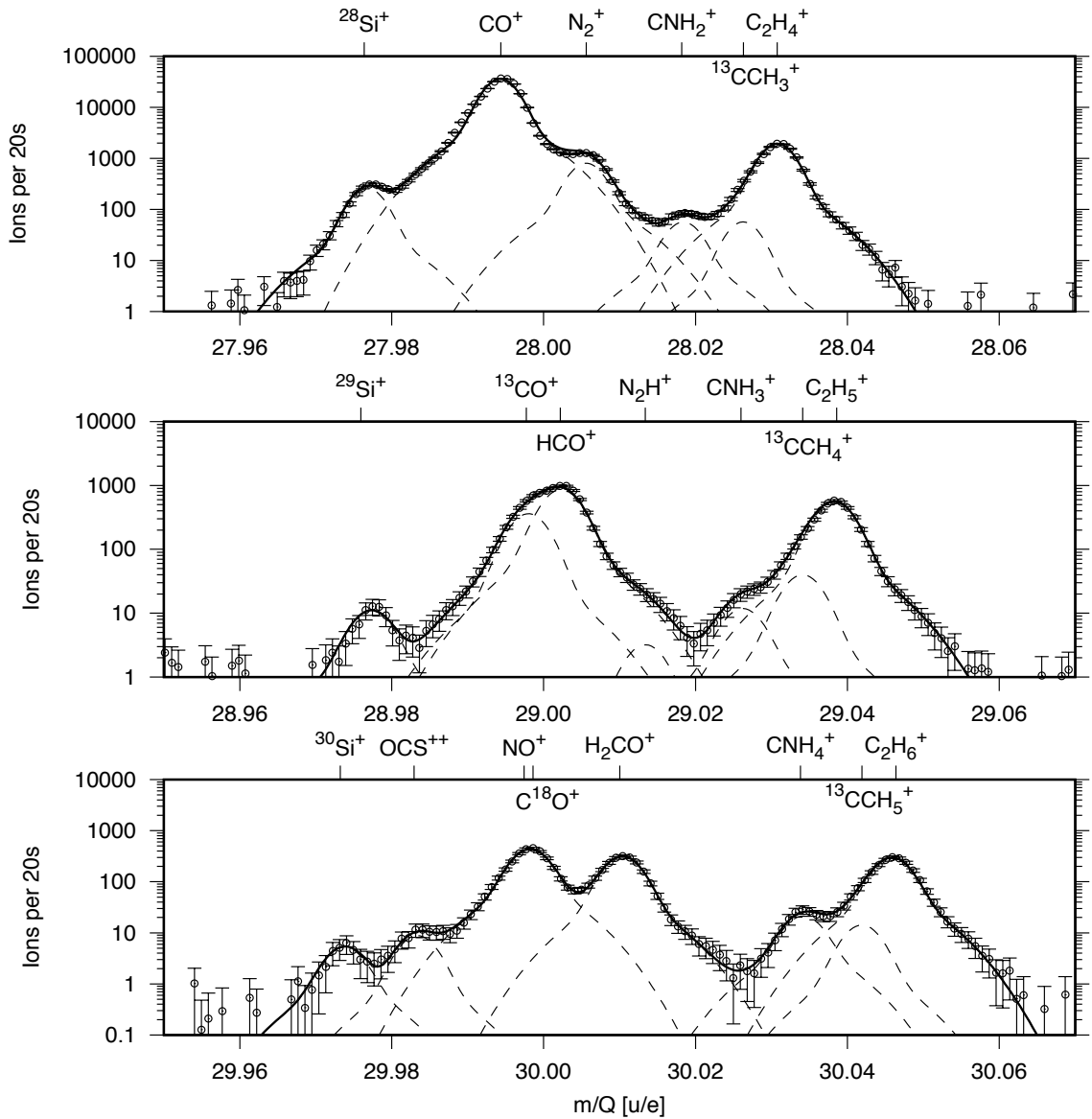}
   \caption{ROSINA DFMS mass spectra on mass-to-charge ratios 28 u/e (top), 29 u/e (middle), and 30 u/e (bottom) obtained during 20~s integration each. The ionized isotopes $^{28}$Si$^+$, $^{29}$Si$^+$, and $^{30}$Si$^+$ have large mass defects and can be found below the corresponding unit mass-to-charge ratio on the left side in the spectra. Among other molecules also carbon monoxide, i.e., the corresponding ions $^{12}$C$^{16}$O$^+$ and $^{13}$C$^{16}$O$^+$ on mass-to-charge ratios 28 u/e and 29 u/e, respectively, can be identified in the exact same data. The dashed lines indicate fitted peaks containing the sum of two Gaussians each, while the black solid line shows the corresponding curve derived from the sum of the individual contributions. The error bars contain counting statistics.}
              \label{FigSpectra}
    \end{figure*}

The ROSINA DFMS is a double focusing mass spectrometer based on the principles of \cite{RN395} in a Nier-Johnson configuration. The DFMS is part of the ROSINA instrument suite \citep{RN13} and is dedicated to the characterization of volatiles and their abundances in the coma of 67P. The DFMS has a high mass resolution of m/$\Delta$m=3000 at the 1\% peak height on the mass-to-charge ratio \deletethree{ yes? I have inserted the word "ratio" throughout where I believe it is needed. If not needed please ignore the suggestions: YES, OK} 28 u/e channel to allow separation of N$_2$ from CO. The DFMS has a field of view $\pm$20$^{\circ}$ for neutral gas and was mounted on Rosetta's instrument platform nominally pointing at the comet.

In the measurement mode used here, the neutral gas particles of cometary origin enter DFMS and are ionized through electron impact inside the ion source. Ions from the coma, however, are efficiently suppressed from being detected by an additional acceleration potential such that their energy exceeds the energy focusing capabilities of the instrument ($\Delta$E/E$\sim$1\%). After extraction from the source and passage through a set of deflection plates, the ion beam is guided through an electrostatic and then a magnetic analyzer before impinging on one of the two rows of the position-sensitive micro-channel plate detector (MCP) mounted along the dispersive direction of the instrument. The electron shower generated in the MCP is then measured on the Linear Detector Array (LEDA). Therefore, even though the DFMS was set to measure neutrals, the particles that were finally detected were the corresponding ions, which included fragments produced during the ionization process inside the ion source.

The potentials applied on the ion optical elements allow only a narrow interval around a chosen integer mass-to-charge ratio to be detected at a time in high resolution (approximately $\pm$~0.25~u/e at 28~u/e). Hence,\deletethree{measurements of different mass-to-charge ions}\changethree{ measurement of ions with different mass-to-charge ratios} are done sequentially.\deletethree{ perhaps: measurement of ions with different mass-to-charge ratios, YES, OK } This leads to the complication that not all species are measured at the same time as it takes roughly 30~s between two different mass-to-charge ratios including approximately 10~s instrument settling time after setting the voltages and 19.8~s integration time. The three Si isotopes on mass-to-charge ratios of 28 u/e, 29 u/e, and 30 u/e were measured back-to-back within 90~s. Given that the measurements are performed sequentially, it is also possible to adjust the MCP voltage (i.e., gain step) individually for each mass-to-charge ratio to amplify low signals while avoiding saturation of the detector by abundant species. This increases the dynamic range of the instrument (10$^{10}$), but requires detailed knowledge of the gain curve of the detector at the different gain steps, especially when deriving isotopic ratios. The DFMS has 16 different gain steps (each amplifying the signal by a factor of $\sim$2.6 over the previous gain step); however, most of the data in the reported measurement period is within one gain step. Another complication comes from the MCP that experienced different charge deposition across the two rows of 512 pixels each. Therefore, individual pixels are in different stages of their aging process, which has to be accounted for. To monitor the aging process, the ion beam is moved across the whole detector during a dedicated calibration campaign and the detector response is recorded. This procedure was repeated at regular intervals throughout the mission and led to different pixel gain curves for various gain steps.

The analysis of the DFMS spectra therefore consists of the following steps: first the detector offset is subtracted, subsequently the measured signal is multiplied by the detector amplification including the aging effects of each individual pixel (gain). Since the foci of this study are the ratios of the three stable Si isotopes, only the relative sensitivities of the isotopes $^{28}$Si, $^{29}$Si, and $^{30}$Si  have to be taken into account. This correction compensates for the mass-dependent fractionation inside the DFMS and contains the decreasing sensitivity for higher mass ions because of slower impact velocities on the detector and thus reduced yield on the MCP. More details on the detector and the data analysis can be found in \cite{RN13} and \cite{RN592}.

\section{DFMS measurements in the coma of 67P/Churyumov-Gerasimenko}\label{section3}

Figure \ref{FigSpectra} shows three example DFMS mass spectra measured in direct sequence at mass-to-charge ratios 28 u/e (top), 29 u/e (middle), and 30 u/e (bottom). The Si isotopes, i.e., the corresponding ions, are found \deletethree{well}\changethree{sufficiently} separated from other molecules to the left of the corresponding integer mass-to-charge ratio as  labeled. Silicon-28 is the most abundant isotope of a refractory element detected by ROSINA at 67P, and it is high enough such that  the two minor stable isotopes $^{29}$Si and $^{30}$Si were also detected. The shapes of the peaks in the DFMS spectra are well known and can be represented by the sum of two Gaussian profiles \citep[see][]{RN505}. The three mass spectra show the contribution from each species as dashed lines and the corresponding envelope (black solid line) fitted to the data. The total amount of detected ions is proportional to the integrated area under the peak, i.e., the corresponding fit in case of overlap resulting from the finite instrument resolution.

   \begin{figure}
   \centering
   \includegraphics[trim = 3.5mm 0mm 0mm 0mm, width=1.02\hsize]{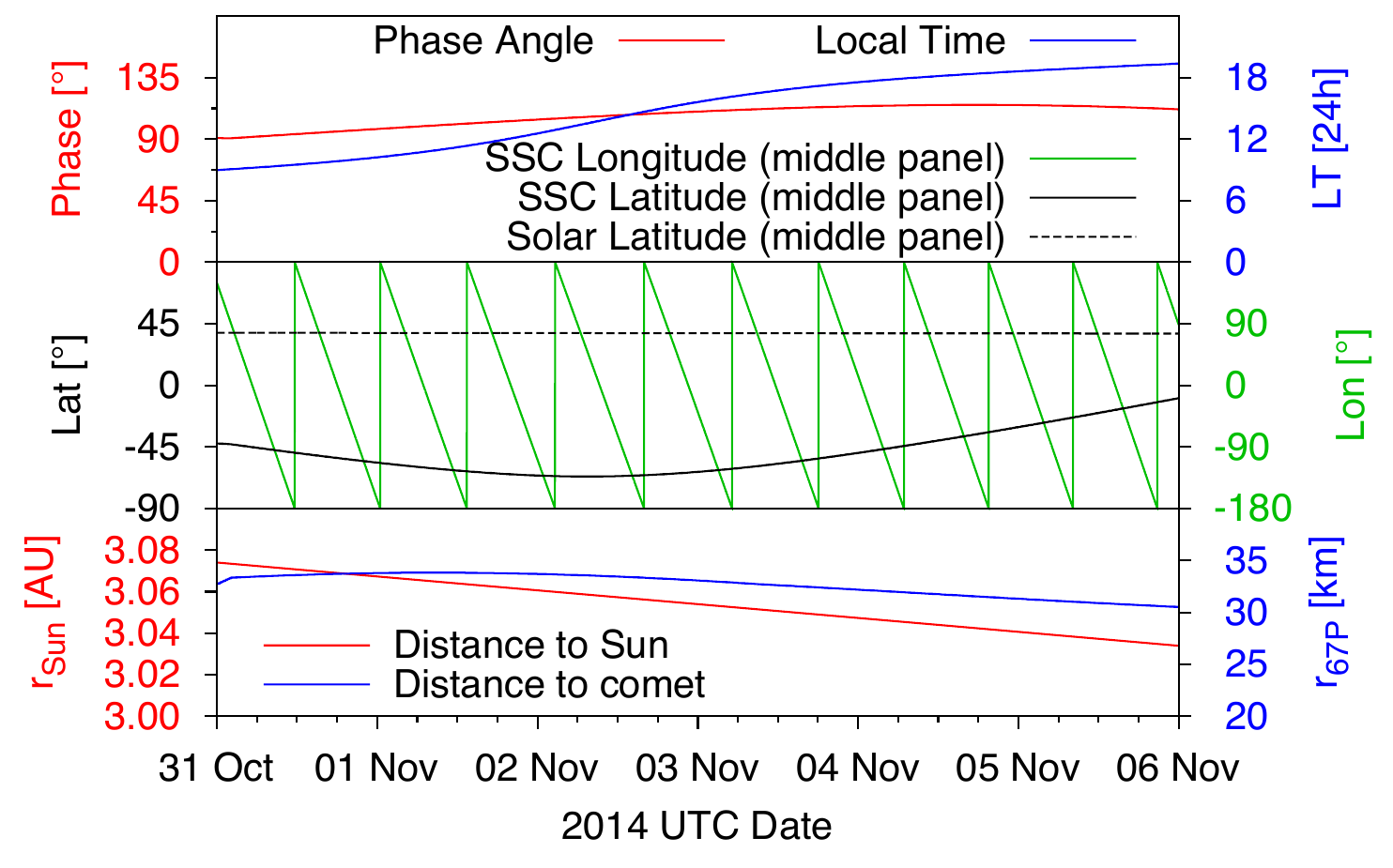}
      \caption{Observation geometry during the period of interest. The top panel shows the phase angle of Rosetta (left y-axis) and the local time of the sub-spacecraft location on a reference sphere (right y-axis). The middle panel shows the subsolar point latitude and the sub-spacecraft latitude (both left y-axes) and longitude (right) on the same reference sphere. The bottom panel shows the heliocentric distance of the comet (left) and the distance of Rosetta from the center of mass of the comet (right). Coordinate system: \cite{Jorda2016257}. }
         \label{FigGeometry}
   \end{figure}
%

A total of 144 mass spectra obtained between 31 October and 6 November 2014 were analyzed for both MCP/LEDA detector rows. Out of 288 mass-to-charge spectra at 28 u/e, 264 were recorded at gain step 15 (second highest amplification). The remaining 16 spectra were recorded at gain step 14. All 288 mass-to-charge spectra at mass-to-charge ratios 29 u/e and 30 u/e were recorded at gain step 16, the highest amplification. Mass-to-charge ratio \deletethree{ same notation should be used throughout : OK} 29 u/e and 30 u/e therefore share the same pixel gain (they are detected on nearly the same pixels on the detector) and overall gain step, and to first order any uncertainty of these cancel when ratios are derived.

Generally the gain steps used were high during the reported period. The reason is that the measurements were performed over the southern hemisphere at less active times  allowing unhindered access for solar wind ions to the comet for subsequent sputtering of material from the surface \citep[see][]{RN561}. Later, when the comet was more active and the solar wind attenuated in the coma, the sputtered signal vanished. Even one of the major species in the coma, carbon monoxide, was relatively low during the observation period. When deriving the ratio at or near the surface of the comet, we also took into account that the density measured at the spacecraft is modified by the different velocities of release. In the sputtering process the velocity difference is proportional to the inverse square root of the mass of the species, $v \sim 1/\sqrt{m}$ \citep{RN561}. \change{The impact of energetic ions onto a solid surface results in the release of atoms through sputtering with a well-defined energy distribution \citep{sigmund1969brief, betz1994energy}. The velocity of the different isotopes hence depends on the mass of the particles, $v = \sqrt{2E/m}$ \changetwo, {and as a consequence, the heavier isotopes move more slowly. Flux conservation results in a higher local density ratio of the heavier to the lighter species measured at Rosetta, e.g., n$_{^{29}\mathrm{Si}}$/n$_{^{28}\mathrm{Si}}$, compared to the ratio of the sputtered fluxes of the same species released from the surface, e.g., f$_{^{29}\mathrm{Si}}$/f$_{^{28}\mathrm{Si}}$:}}

\begin{displaymath}
\indent \mathrm{\frac{n_{^{29}Si}}{n_{^{28}Si}}=\frac{f_{^{29}Si}}{f_{^{28}Si}}\frac{v_{^{28}Si}}{v_{^{29}Si}}=\frac{f_{^{29}Si}}{f_{^{28}Si}}\sqrt{\frac{\mathrm{m_{^{29}Si}}}{\mathrm{m_{^{28}Si}}}}\approx 1.018~\frac{f_{^{29}Si}}{f_{^{28}Si}}}
\end{displaymath}

Figure \ref{FigGeometry} shows that the observing conditions remained similar during the measurement period. Rosetta was just outside of 30 km from the center of mass of the comet (bottom panel). The heliocentric distance changed only marginally and variations in the flux of solar protons hitting the cometary surface are associated with variations of the solar wind. Because the three isotopes were measured sequentially, the short-term variations in the solar wind flux on timescales of minutes affect the derived ratios and further increase the scatter \deletethree{on top of, OK} \changethree{in addition to} the scatter due to low count statistics. However, 144 measurements for each of the two MCP rows distributed over the course of 6 days will even out such short-term variations. We therefore co-added all mass spectra of the same mass-to-charge ratio from which the results were derived, and for comparison averaged the 288 individual ratios (weighted by their errors), which leads to the same conclusions. The middle panel in Figure \ref{FigGeometry} shows the sub-spacecraft (SSC) longitude and latitude.  The subsolar latitude is also indicated and shows that the northern hemisphere was mostly illuminated and active \citep{RN593} while the spacecraft moved above the southern hemisphere at local winter and covered almost 12 cometary rotations. Rosetta flew close to the terminator plane (or slightly on the night side) and on the afternoon side indicated by the phase angle and local time in the top panel, respectively.

Given the large angular acceptance of DFMS and the distance of $\sim$30 km, the whole comet was in the field of view and the sputtered species originated from almost anywhere on the nucleus when exposed to solar wind. The results are thus different from what is typically seen, for example from the analysis of  presolar SiC, graphite, and Si$_3$N$_4$ inclusions, which are large enough to be analyzed individually. Here each measurement represents an average over the part of the surface in the field of view and accessible to solar wind.
The results are reported in the $\delta$-notation together with 1-$\sigma$ errors (and appended are the ratios of the heavy to the light isotope and the ratio of the light to the heavy isotope):

\begin{displaymath}
\indent \mathrm{\delta^{29}Si=1000\times\left(\frac{^{29}Si/^{28}Si\big |_{67P}}{^{29}Si/^{28}Si\big |_{solar}}-1\right),}
\end{displaymath}

\begin{displaymath}
\indent \mathrm{\delta^{30}Si=1000\times\left(\frac{^{30}Si/^{28}Si\big |_{67P}}{^{30}Si/^{28}Si\big |_{solar}}-1\right),}
\end{displaymath}

\noindent and 

\begin{displaymath}
\indent \mathrm{\delta^{30}Si/^{29}Si=1000\times\left(\frac{^{30}Si/^{29}Si\big |_{67P}}{^{30}Si/^{29}Si\big |_{solar}}-1\right),}
\end{displaymath}

\noindent which gives the observed deviations of cometary material compared to solar material in permil ($\permil$). Uncertainties were derived by error propagation taking into account 10\% uncertainty of the relative detector gain amplification (3\% for $\delta^{30}$Si/$^{29}$Si as this ratio is measured on the same gain step), an uncertainty of 5\% due to variation in the solar wind proton flux remaining after the averaging process per mass-to-charge difference (10\% between 28 u/e and 30 u/e, and 5\% between 28 u/e and 29 u/e and 29 u/e and 30 u/e), and count statistics.

Combining all 288 measurements shows a depletion of the heavy  isotopes $^{29}$Si and $^{30}$Si relative to  $^{28}$Si and solar abundances. Summing up all mass spectra yields for $^{29}$Si with respect to $^{28}$Si:
\begin{align*}
\indent \indent \mathrm{\delta^{29}Si}&=(-145\pm98)\permil \\
\mathrm{^{29}Si/^{28}Si}&=0.0434\pm0.0050 \\
\mathrm{^{28}Si/^{29}Si}&=23.0\pm2.6
\end{align*}

\noindent for $^{30}$Si with respect to $^{28}$Si:
\begin{align*}
\indent \indent \mathrm{\delta^{30}Si}&=(-214\pm115)\permil \\
\mathrm{^{30}Si/^{28}Si}&=0.0263\pm0.0038\\
\mathrm{^{28}Si/^{30}Si}&=38.0\pm5.6
\end{align*}

\noindent and correspondingly:
\begin{align*}
\indent \mathrm{\delta^{30}Si/^{29}Si}&=(-97\pm55)\permil \\
\mathrm{^{30}Si/^{29}Si}&=0.595\pm0.036 \\
\mathrm{^{29}Si/^{30}Si}&=1.7\pm0.1 
\end{align*}

\begin{figure*}
\centering
\includegraphics[width=150mm]{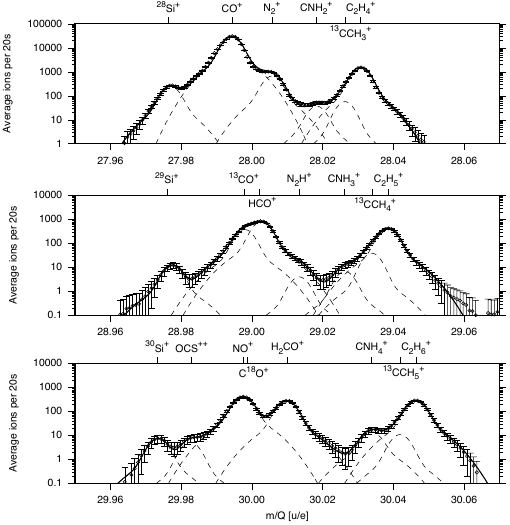}
\caption{
   Same as Figure \ref{FigSpectra}, but co-added over all 288 individual mass spectra and used to derive the reported isotopic ratios of $^{29}$Si/$^{28}$Si, $^{30}$Si/$^{28}$Si, $^{30}$Si/$^{29}$Si, and $^{13}$C/$^{12}$C in CO, C$_2$H$_4$, and C$_2$H$_5$.\deletethree{, respectively. see there are 4 isotope ratios and 3 compounds, so  respectively does not work. Perhaps you mean "from top to bottom" ?  YES, JUST REMOVE RESPECTIVELY}}
\label{FigSpectraAverage}
\end{figure*}

Uncertainties in the relative signal amplification between the different mass-to-charge spectra affect the obtained isotopic ratios. Fortunately, there is another isotopic ratio in the exact same dataset that can be used for comparison. The top panel in Figure \ref{FigSpectra} shows the presence of CO and the middle panel the corresponding molecule with the heavy C isotope, $^{13}$C. For reference we used the abundances of 98.892\% for $^{12}$C  and 1.108\% for $^{13}$C \citep[$^{13}$C/$^{12}$C=0.0112,][]{RN664}, again within the ranges provided by the Commission on Isotopic Abundances and Atomic Weights of the International Union of Pure and Applied Chemistry \citep{berglund2011isotopic}. Compared to these values the solar wind appears to be depleted in $^{13}$C by approximately 10\% as shown by ion probe isotopic measurements of carbon trapped in the top 50 nm of the lunar regolith \citep{0004-637X-600-1-480}.

Carbon monoxide is one of the dominant cometary volatiles at 67P at that time of the mission \citep{RN592} even though the absolute amount outgassing from the winter hemisphere is slightly reduced with respect to summer \citep{RN476}. Still, the signal is much stronger compared to Si. However, an extra difficulty arises, due to the closeness of the peaks of $^{13}$CO and HCO, which are of similar mass and therefore add uncertainty to the fitted area under the peak. The CO signal shown here contains contributions from parent CO in the coma and a negligible contribution of CO from spacecraft background \citep{RN384}. Furthermore, CO is  formed through electron impact dissociation (e.g., of CO$_2$)\deletethree{ok? to avoid a punctuation problem, YES, OK} inside the DFMS ion source \citep{RN592}, which amounts to 10~--~20\% of the total CO signal during the investigated period. We obtain a mean 
\begin{align*}
\indent \indent \mathrm{\delta^{13}C}&=(34\pm103)\permil \\
\mathrm{^{13}C/^{12}C}&=0.0116\pm0.0012 \\
\mathrm{^{12}C/^{13}C}&=86.0\pm8.5
\end{align*}
\noindent in CO, consistent with bulk planetary and meteoritic C abundances; the error bar contains 10\% pixel gain and fit uncertainty and a minor contribution from counting statistics. This is supported by the detailed analysis of the 67P C isotope ratio $\delta^{13}$C=60$\pm$50 in CO$_2$ \citep[$^{13}$C/$^{12}$C=0.0119$\pm$0.0006, $^{12}$C/$^{13}$C=84$\pm$4;][]{haessig}. Inside the DFMS the CO signal has different contributions: CO$_2$ is fragmented inside the ion source which amounts to 10-20\% of the CO signal; however, we did not separate the individual contributions to the CO signal as the observed $^{13}$C/$^{12}$C ratio is much closer \changethree{the \cite{haessig} values and} to bulk planetary and meteoritic C \deletethree{and the H\"assig et al. (submitted) values compared} \changethree {in comparison} to the deviation observed in the Si isotopes.\deletethree{ I had to avoid a punctuation problem here too. ok like so? YES, OK}

Figure \ref{FigSpectraAverage} shows the co-added mass spectra of all 288 individual measurements per mass-to-charge ratio. This has been used to investigate the $^{13}$C/$^{12}$C isotopic ratio in two other molecules in the exact same dataset, namely in C$_2$H$_4$ and C$_2$H$_5$, which are  both  most likely daughter products of ethane, C$_2$H$_6$ \citep{RN592}. The former yields 
\begin{align*}
\indent \indent \mathrm{\delta^{13}C}&=(65\pm151)\permil \\
\mathrm{^{13}C/^{12}C}&=0.0120\pm0.0017 \\
\mathrm{^{12}C/^{13}C}&=83.6\pm11.8
\end{align*}
\noindent in C$_2$H$_4$ and includes 10\% detector gain uncertainty, 10\% fitting uncertainty, and a minor contribution from count statistics error. The latter yields
\begin{align*}
\indent \indent \mathrm{\delta^{13}C}&=(41\pm109)\permil \\
\mathrm{^{13}C/^{12}C}&=0.0117\pm0.0012 \\
\mathrm{^{12}C/^{13}C}&=85.5\pm9.0
\end{align*}
\noindent in C$_2$H$_5$ and includes 3\% detector gain uncertainty (same gain step thus smaller error due to different peak centroids, see Figure \ref{FigSpectraAverage}), 10\% fitting uncertainty, and a minor contribution from count statistics.

Unfortunately the resolving power of the DFMS is not high enough to separate C$^{18}$O in the shoulder of NO, and therefore the $^{18}$O/ $^{16}$O isotopic ratio in CO cannot be investigated. Obviously the investigated period is not particularly well suited to deriving C isotopic ratios and this unsuitability is reflected in the error bars. We therefore refer to the work by \cite{haessig} who \change{chose periods specifically for C and O isotopes}, which, however, were not good for the detection of the three Si isotopes. Nevertheless, the combination of all these measurements indicates that possible mass dependent fractionation processes inside the instrument, including the applied electron-impact ionization technique, cannot be the cause of the measured depletion of the heavy Si isotopes.

\begin{figure*}
   \centering \includegraphics[width=185mm]{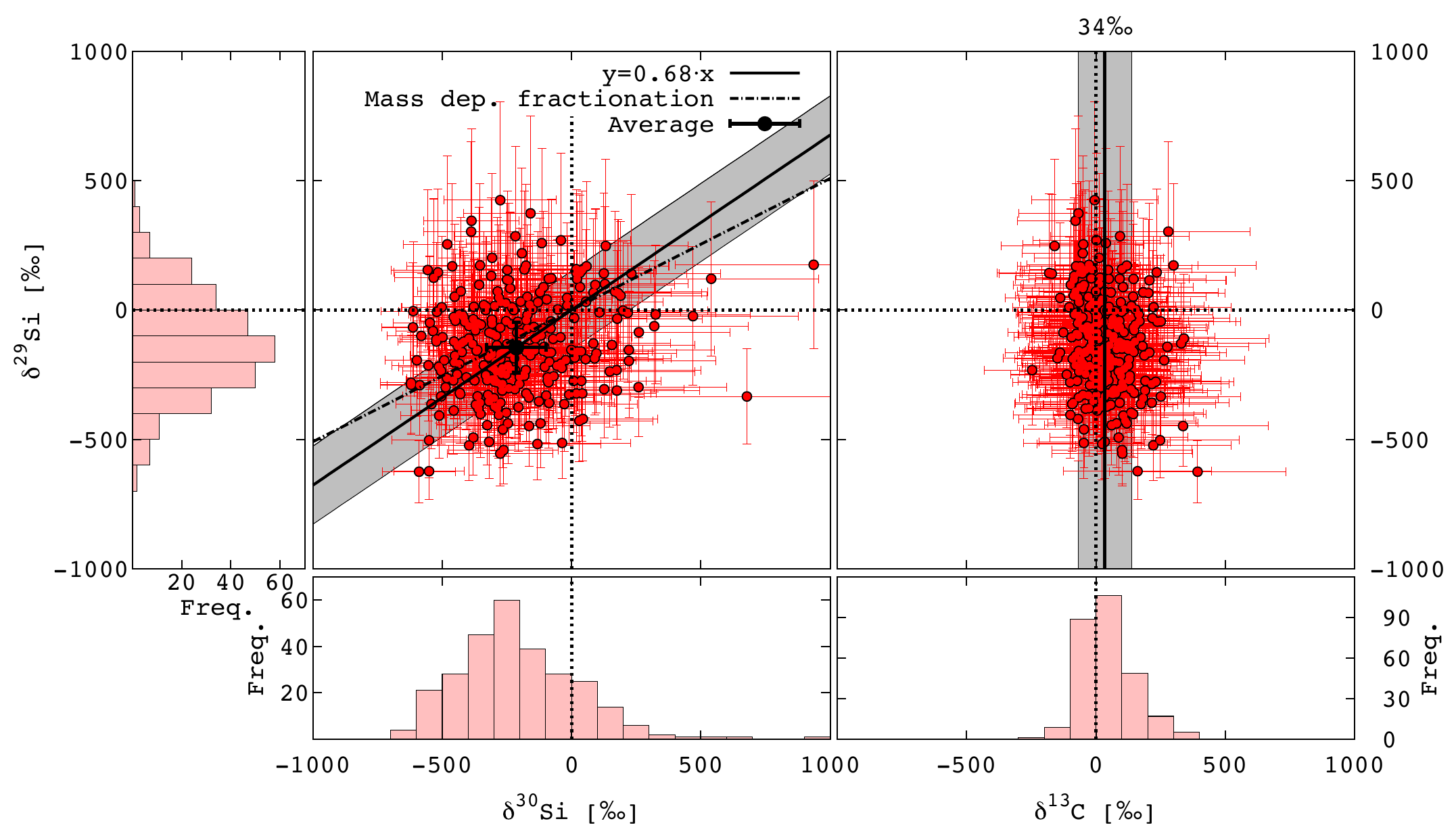}
   \caption{Measured isotopic ratios of Si plotted in $\delta$-notation depicting deviations in permil ($\permil$). Left: Individual $\delta^{29}$Si vs. $\delta^{30}$Si ratios (red points) and average (black point). Also shown is the expected line from mass dependent fractionation (black dash-dotted line) and a line through the origin and average (black solid line, slope $\sim$0.68) with error (gray area, see text). Right:  Data points showing individual $\delta^{29}$Si vs. $\delta^{13}$C in carbon monoxide and the black solid line the average $\delta^{13}$C~=~34~$\permil$ with error (gray area, see text). The black dashed lines indicate the corresponding reference values in both plates from \cite{RN664}. Below and on the left side are the histograms of the individual measurements.}
              \label{FigDeltaPlot}
    \end{figure*}

It is also possible to investigate individual ratios derived from three sequentially measured single spectra as represented in Figure \ref{FigSpectra}. The left panel in Figure \ref{FigDeltaPlot} shows the ratios of $^{28}$Si, $^{29}$Si, and $^{30}$Si in $\delta$-notation. When the ratios are solar they fall on the dashed lines, while the observed deviations at 67P from solar are given in permil ($\permil$). The red points show the individual ratios, derived from three subsequent measurements of mass-to-charge ratios 28 u/e, 29 u/e, and 30 u/e within 90~s.
For the individual ratios the sputtered signal is low, in particular for the minor isotopes $^{29}$Si and $^{30}$Si, which leads to large error bars (including a  10\% error due varying solar wind conditions). The black point denotes the ratio derived from the averaged spectra shown in Figure \ref{FigSpectraAverage}.

To further investigate possible isotope fractionation processes a connecting (solid black) line is drawn between the origin and the average of all points (the resulting slope is $\sim$0.68). The gray envelope shows the error of the average, 
\begin{displaymath}
\indent \mathrm{\sigma=\sqrt{\left(\sigma_{\mathrm{\delta^{29}Si}}\right)^2+\left(\mathrm{\sigma_{\delta^{30}Si}}\right)^2}}.
\end{displaymath}
\noindent For reference, the mass dependent  fractionation line (black dash-dotted) through the origin with slope
\begin{displaymath}
\indent \mathrm{\beta=\left. ln\left(\dfrac{m_{^{28}Si}}{m_{^{29}Si}}\right)\:\middle/\,ln\left(\dfrac{m_{^{28}Si}}{m_{^{30}Si}}\right) \right.=0.5092}
\end{displaymath}
\noindent is added (see \citealt{RN670}). The discussion  follows in the next section.

The right panel of Figure \ref{FigDeltaPlot} shows $\delta^{29}$Si versus $\delta^{13}$C in CO with the black solid line denoting the average isotope composition. Despite the difficulty  distinguishing $^{13}$CO from HCO, the horizontal spread (indicated by the gray area with $\sigma_{\delta^{13}C}$ error) is smaller because the  count rates are higher and because the ratio is not sensitive to variations in the solar wind.

\section{Discussion and conclusions}
Strongly depleted heavy Si isotopes with respect to solar compositions are rare in solar system materials \citep{nguyen2010coordinated, nguyen2014resolving, floss2016presolar}. The Si isotope composition of the majority of the matter is either close to solar values or exhibits a minor enhancement of heavy isotopes. The ROSINA DFMS, however, indicates a  depletion in the heavy isotopes $^{29}$Si and $^{30}$Si at comet 67P, which are different from the solar abundances by 1-$\sigma$ to 2-$\sigma$. \change{In the following discussion we  focus on the possible reasons for this depletion, bearing in mind that our results do not fully exclude solar  Si isotope abundances.}

A possible explanation is the fractionation of solar system material. The scatter of the individual observations shown in Figure \ref{FigDeltaPlot} is considerable, and therefore our results are also consistent with mass dependent fractionation. Material in primitive meteorites including refractory inclusions such as CAIs are to some degree homogenized and reprocessed in the solar system, unlike interstellar silicates, graphite, diamond, and silicon carbide, which constitute presolar material \citep{floss2016presolar}. Therefore, the measured deviations from solar isotopic ratios in most meteoritic material are much smaller than the variations of Si isotopes measured in presolar  grains and are inconsistent with ROSINA measurements. This is also true for  equilibrium Si isotope fractionation between gaseous SiO and solid forsterite during condensation from nebular gas that yields much smaller degrees of fractionation \citep{RN677}. Our results do not exclude additional mass-dependent fractionation, which can occur in equilibrium or in kinetically controlled physical or chemical processes such as diffusion, evaporation, and phase changes \citep{RN655,Chakrabarti20106921,Zambardi201367}. It is clear, however, that the measured depletion is large for any kind of mass dependent fractionation and different from most of the material found in the inner solar system. Furthermore, earlier measurements by ROSINA \citep{RN561} and COSIMA \citep{RN623} independently confirm increased Si abundances compared to solar \change{abundances}. This would be compatible with a silicate enrichment of the refractory material.

Further possibilities are solar and cosmic irradiation of the exposed surface material that could induce mass dependent isotopic fractionation\delete{ processes}. At the time of the observations, Rosetta was above the less illuminated southern hemisphere. However, this part of the nucleus became much more active as the comet approached perihelion and the subsolar point moved to the southern hemisphere. It is estimated that the comet lost several meters of surface material \delete{during one}\change{per} apparition \citep{RN616,Hansen26092016} and thus exposed new and fresh material while shedding the old. Similar to the case of the abundant O$_2$ found in the comet, which cannot be explained by radiation induced processes \citep{RN665,RN668}, it is unlikely that the surface material received high enough doses of radiation during the 6.4~yr orbital period of the comet to explain the observed degree of fractionation.

On the other hand, we assume no elemental fractionation took place during the sputtering process on the comet's surface. Sputtering releases material from the surface in approximately stoichiometric proportions into the exosphere after a steady state of the sputter process has been established \citep{RN672}, and the composition of the flux of sputtered atoms will reflect the average bulk composition of the particles. Reaching this steady state takes about 1~month in the solar wind for a freshly exposed grain \citep{RN673} on Mercury\delete{ which is still short compared to 67P's orbital period}. On 67P dust is transported around the nucleus, especially from the southern to the northern hemisphere when the comet is close to the Sun. \change{The southern hemisphere, above which the Si isotopes were measured, was inactive for several years prior to the measurements and the material was left exposed on the surface since the last apparition. Nevertheless, the transport of grains from north to south cannot be fully discarded, and potentially exposed fresh material releases material with Si isotope ratios different from bulk upon sputtering by solar wind protons.} When dust is freshly exposed altered layers can form that are deeper than the topmost two to three atomic layers from which the sputtered products originate. Non-steady-state situations, possibly favoring either amorphous or crystalline forms of Si can potentially alter the measured elemental abundances but less so the isotopic ratios. Also, at low temperatures, i.e., with negligible thermal diffusion, steady state situations can be reached where the sputtered abundances correspond to the bulk concentrations \citep{RN672}. \change{\changethree{Nevertheless, in Secondary Ion Mass Spectrometry (}SIMS\changethree{)}, \deletethree{ abbreviation to introduce? YES} deviations of the sputtered fluxes compared to bulk abundances can also be observed  for isotopes. Hence, reference samples are used for calibration. Our measurement setup, with the primary ion beam consisting of solar wind protons and the sputtering process which occurs on the surface of the comet, adds an extra level of complexity that we cannot fully account for. \citet{shimizu1982isotope} report enriched lighter isotopes up to a few percent in the released secondary ions. The effect is approximately proportional to the ratio of the involved masses and might in part explain the observations, even though the  DFMS measured the neutrals from the sputtering process as opposed to the ions in SIMS.}

\change{After leaving the surface the isotopes \deletetwo{to} have different velocities, due to their different masses (see Section \ref{section3}). The correction we applied to the local density ratio to obtain the sputtered flux ratio is on the order of a few percent, e.g.,  $\sqrt{m_{^{28}Si}/m_{^{29}Si}} = 0.983$ for the $^{29}$Si/$^{28}$Si ratio,  and is minor compared to the observed deviation. Once inside the instrument we use the lab derived calibration to correct for mass dependent fractionation inside the sensor and compare our results to the C isotope ratios in the volatiles for reference (Fig.~\ref{FigDeltaPlot}).}

The measured depletion of the heavy Si isotopes could also point to a presolar origin of the Si in 67P that has not been homogenized and reprocessed (even an extrasolar origin cannot be ruled out). This would be surprising considering that a significant fraction of the \deletethree{stardust} minerals \changethree{found by the Stardust mission to comet 81P/Wild} seem to originate from the inner solar system \citep{mckeegan2006isotopic,brownlee2014stardust}. On the other hand, at least parts of the cometary ices in 67P seem to have survived the formation of the solar system \citep{Calmonte08102016,AltweggD2O} and in the case of water have inherited D/H ratios similar to ISM material. If this is true,  the refractory material enclosed in such icy grains should also be preserved, which in turn hints at different formation scenarios for Wild 2 and 67P. There are only a few measurements of ISM Si isotope ratios;  e.g., \cite{RN680} report solar Si isotope ratios in the galactic center region and thus suggest no galactic gradients in the $^{29}$Si/$^{30}$Si ratio (see also \citealt{RN679}).  More recently, however, \cite{RN678} report depleted heavy Si isotopes in the Orion KL region, albeit with large error bars: $\delta^{29}$Si = $(-242\pm291)\permil$ ($^{29}$Si/$^{28}$Si = 0.0385$\pm$0.0148, $^{28}$Si/$^{29}$Si = 26$\pm$10) and $\delta^{30}$Si/$^{29}$Si = $(-108\pm315)\permil$ ($^{30}$Si/$^{29}$Si = 0.588$\pm$0.208, $^{29}$Si/$^{30}$Si = 1.7$\pm$0.6).

The most abundant Si-bearing presolar grains are silicates that are mostly identified by their anomalous O isotopic compositions \citep{messenger2003samples}. For grains showing anomalous O isotopic compositions, \citet{nittler1997stellar} proposed a classification scheme of four groups based on the likely  stellar origin. \deletethree{ have I interpreted correctly? } Groups 1 to 3 could originate from asymptotic giant branch (AGB) stars of different metallicities and at different evolution stages, whereas the origin of group 4 grains is still unclear \citep{hoppe2009stardust, floss2016presolar}. Silicon isotopes have been measured for a number of presolar grains, whether they are anomalous in O isotopes or not. From the available presolar grain data (presolar grain database, \citealt{hynes2009presolar}), the few presolar silicates showing Si isotopic composition compatible with the average Si isotopic composition of 67P would mostly belong to group 1 presolar silicates. It should be noted that this could reflect a statistical effect, as type 1 presolar grains are about 10 times more abundant than those from groups 2 to 4. \cite{nguyen2014resolving} also pointed out possible dilution effects due to the small size of presolar grain compared to the size of the analyzing beam during NanoSIMS measurements. The range of variation of Si isotopic data for group 1 grains compared to those in groups 2 to 4 shows both heavy Si isotope enrichments and depletions. Group 1 presolar grains show O isotopic compositions enriched in $^{17}$O and about solar (to subsolar) $^{18}$O/$^{16}$O. They are thought to originate from O-rich low-mass red giant and AGB stars, with an almost solar metallicity \citep{nittler1997stellar}.

Supernova (SN) silicates have been found in a few objects, mostly based on O isotopic measurements \citep{messenger2003samples,bland2007cornucopia,floss2009auger}. When measured, the Si isotopes of these SN grains usually have close to normal $\mathrm{\delta^{30}Si}$ and negative $\mathrm{\delta^{29}Si}$ values, but a dilution effect with normal surrounding silicate could affect the measurements (as mentioned by \citealt{nguyen2014resolving}). Supernova silicates show large $^{18}$O enrichments and $^{17}$O depletions. Unfortunately, the O isotopes of these grains could not be determined by ROSINA as the signal is dominated by O from volatiles.

Less abundant Si-bearing presolar grains are SiC, which are 5 to 10 times less abundant than presolar silicates. Silicon isotopes of mainstream presolar SiC show enrichment in the heavy Si isotopes up to 200~$\permil$ and a slope of 1.35 in \change{a} three-isotope plot, reflecting the variation in initial parent stellar composition and galactic cosmic evolution, as well as possible local heterogeneities in the local interstellar medium. Galactic cosmic evolution predicts that the $^{29}$Si/$^{28}$Si and $^{30}$Si/$^{28}$Si ratios increase with galactic age \citep{timmes1996galactic}. During the AGB phase, slight enrichments in heavy Si isotopes can be observed, but they will not become significant until the star becomes C-rich. So the Si isotopic compositions of silicate grains should reflect the parent star metallicity. This is observed as the Si isotopic signature of presolar silicates are systematically depleted in $^{29}$Si and $^{30}$Si compared to that of mainstream SiC. SiC-X grains are thought to originate from supernovae and show the \delete{opposite}\change{complementary} isotopic signature of mainstream SiC. However, they constitute only roughly 1\% of all meteoritic SiC \citep{RN656}. X-grains are depleted in heavy Si isotopes \citep{RN659} with a slope slightly steeper than expected from mass dependent fractionation (slope 0.6~--~0.7; see \citealt{RN656}). SiC-X grains are thought to show signatures of advanced nuclear burning stages and inhomogeneous mixing of matter during the explosion of the supernova; however, some observations such as the enrichment of $^{29}$Si with respect to $^{30}$Si and the excess of $^{15}$N seem to be in contradiction with this interpretation \citep{RN656}. SiC-X grains also show peculiar isotopic ratios in other elements such as N, C, and Al. None of these species is accessible to ROSINA either, due to a strongly dominating contribution from the volatile phase of the comet (e.g., the C-signal in DFMS is dominated by fragments from the major volatiles CO and CO$_2$  broken up in the electron-impact ionization process inside the ion source) or sputtered fluxes below the detection limit (Al). Other types of interstellar grains containing Si and showing depletion in the heavy isotopes are silicon nitride grains (Si$_3$N$_4$, slope 0.67; \citealt{RN660}) and in graphite \citep{RN661,RN662,RN663}. However, similar to the SiC-X grains, these grain types  seem to be very rare.

Comets are chemically very heterogeneous as indicated by their variation in the measured D/H  in water \citep{RN462}. Earlier observations of 67P show highly volatile species and isotopic ratios corresponding to ices formed in the ISM. If this is also the case for part of the refractory material in 67P, this would in turn limit the amount of processed presolar material in the comet forming region and limit the radial mixing of matter during the formation of our solar system.

\change{Presolar matter makes up only a tiny fraction of primitive materials and has very little influence on the bulk isotopic ratio. However, as} \cite{RN655} pointed out, there is a mismatch between the average isotopic ratios of presolar grains known to date and the average solar ratios. If we investigate the heavy isotopes of Si, $^{29}$Si, and $^{30}$Si, we find that, except for a small fraction, most known presolar matter has  solar Si isotope composition or is enriched in the heavy Si isotopes and thus most likely represents only one part of the material from which the solar system formed.

Our analyses show that the heavy Si isotopes at 67P are depleted. 67P most likely is not representative of all comets; it is  instead one of a subset that formed far from the Sun. Its pristine composition makes it very relevant to the study of the matter forming our solar system. We therefore expect comets to add crucial information on the initial composition and formation scenarios. Complementary measurements by Rosetta COSIMA and possible future comet sample return missions are indispensable in this investigation.

\begin{acknowledgements}
 ROSINA would not have given such outstanding results without the work of the many engineers, technicians, and scientists involved in the mission, in the Rosetta spacecraft, and in the ROSINA instrument team over the last 20 years, whose contributions are gratefully acknowledged. Rosetta is a European Space Agency (ESA) mission with contributions from its member states and NASA. We acknowledge here the work of the whole ESA Rosetta team. Funding: Work at University of Bern was funded by the State of Bern, the Swiss National Science Foundation, and the ESA PRODEX (PROgramme de D\'eveloppement d'Exp\'eriences scientifiques) program. Work at Southwest Research Institute was supported by subcontract \#1496541 from the Jet Propulsion Laboratory (JPL). Work at the Royal Belgian Institute for Space Aeronomy (BIRA-IASB) was supported by the Belgian Science Policy Office via PRODEX/ROSINA PRODEX Experiment Arrangement 90020. This work was supported by CNES (Centre National d'Etudes Spatiales). Work at the University of Michigan was funded by NASA under contract JPL-1266313.\newline
 
Data and materials availability: All ROSINA data have been and will be released to the Planetary Science Archive of ESA (http://www.cosmos.esa.int/web/psa/rosetta) and to the Planetary Data System archive of NASA (https://pds.nasa.gov/). All data needed to evaluate the conclusions in the paper are present in the paper. Additional data related to this paper may be requested from the authors.
\end{acknowledgements}

%
%

\end{document}